\documentclass[letterpaper, 10 pt, conference]{ieeeconf}  
\IEEEoverridecommandlockouts                              
\usepackage{mathtools}
\usepackage{amsmath}
\usepackage{amssymb}
\usepackage{mathrsfs}
\usepackage{graphicx}
\usepackage{subfigure}
\usepackage{bm}
\usepackage{hyperref}
\usepackage{enumerate}

\allowdisplaybreaks[4]

\newtheorem{asmp}{\textbf{Assumption}}

\newtheorem{thm}{\textbf{Theorem}}
\newtheorem{exmp}{\textbf{Example}}
\newtheorem{rmk}{\textbf{Remark}}

\newcommand{\mtcc}{\mathcal{C}}

\newcommand{\mtce}{\mathcal{E}}

\newcommand{\mtcg}{\mathcal{G}}

\newcommand{\mtcs}{\mathcal{S}}

\newcommand{\mtcv}{\mathcal{V}}

\newcommand{\bfl}{\mathbf{1}}

\newcommand{\bfz}{\mathbf{z}}

\newcommand{\RR}{\mathbb{R}}
\newcommand{\EE}{\mathbb{E}}

\newcommand{\NN}{\mathbb{N}}

\newcommand{\Let}{: =}

\newcommand{\sgn}{\textup{sgn}}

\title{\LARGE \bf
What is the Expected Transient Behavior of \\Opinion Evolution for Two Communities?
}

\author{Yu Xing  and Karl H. Johansson 
\thanks{This work was supported by the Knut \& Alice Wallenberg Foundation and the Swedish Research Council. The authors are with Division of Decision and Control Systems, School of Electrical Engineering and Computer Science, KTH Royal Institute of Technology, and with Digital Futures, Stockholm, Sweden. Email:
        {\tt\small \{yuxing2,kallej\}@kth.se}}
}

\begin{document}

\maketitle

\begin{abstract}
	We study the transient behavior of a gossip model, in which agents randomly interact pairwise over a weighted graph with two communities. Edges within each community have identical weights, different from the weights between communities. It is shown that, at the early stage of the opinion evolution, the expected agent states in the same community have identical sign, despite influence of stubborn agents. Moreover, it is shown that the expected states of the agents in the same community concentrate around the initial average opinion of that community, if the weights within communities are larger than between. In contrast, if the edge weights between communities are larger, then the expected states of all agents concentrate around everyone's initial average opinion. Different from the traditional asymptotic analysis in the opinion dynamics literature, these results focus on the initial phase of opinion evolution and establish a correspondence between community structure and transient behavior of the gossip model. The results are illustrated by numerical examples.
\end{abstract}

\section{Introduction}\label{sec_intro}

Opinion dynamics is a field that studies how interpersonal influence shapes individual opinions. In recent years, it has been investigated extensively in the control community. Most of the research has focused on convergence and stability analysis~\cite{proskurnikov2017tutorial,proskurnikov2018tutorial}. Less attention has been paid to the analysis of transient behavior of opinion dynamics, in contrast to asymptotic behavior. In reality it could be hard to tell whether a social network reaches its steady state or not, and whether its communities behave relatively uniformly during the initial phase. It is known that community structure~\cite{fortunato2016community} plays a key role in group dynamics. Therefore, there is a need to establish quantitative correspondences between the community structure and behavior of the opinion formation process. In practice, such results can also inspire the design of community detection algorithms based on state observations~\cite{schaub2020blind,xing2021detecting} and the design of model reduction tools for large-scale networks~\cite{cheng2018model,niazi2020state}.

\subsection{Related Work}
There are at least three broad classes of models explaining consensus and clustering phenomena in social opinions~\cite{proskurnikov2018tutorial,flache2017models}: models of assimilative, homophily, and negative social influences. In the first class of models, agents update their opinions according to the average of their neighbors' opinions. A seminal example is the DeGroot model~\cite{degroot1974reaching}. The Friedkin--Johnsen (FJ) model~\cite{friedkin1990social} generalizes the DeGroot model by assuming that agents can be stubborn with respect to their initial belief. The second class of models, which includes the Hegselmann--Krause (HK) model~\cite{hegselmann2002opinion} and the Deffuant--Weisbuch (DW) model \cite{deffuant2000mixing}, assumes that only individuals holding similar opinions can interact with each other. These models are prone to generate clustering phenomena. The third class of models allows negative edges that can increase belief difference between agents, and often ends in polarization~\cite{proskurnikov2018tutorial,shi2019dynamics}. 

This paper considers transient behavior of a gossip model with stubborn agents. This model is a stochastic counterpart of the DeGroot model. Consensus of the gossip model has been extensively studied~\cite{boyd2006randomized,fagnani2008randomized}. The authors in~\cite{acemouglu2013opinion} show that
the existence of stubborn agents may explain the fluctuation of social opinions. 
Less attention has been paid to transient behavior of group dynamics, in contrast to asymptotic analysis. The authors in~\cite{banisch2012agent} investigate the transient stage of discrete-state Markov chains. The paper~\cite{barbillon2015network} studies quasi-stationary distributions of a contact process, and~\cite{xiong2017modeling} analyzes transient opinion profiles of a voter model. The authors in~\cite{dietrich2016transient} provide conditions for detecting transient clusters in a generalized HK model.

The study of community and community detection~\cite{fortunato2016community} has a long history. Early research defines communities as complete subgraphs in a network~\cite{festinger1949analysis}. One popular modern definition of communities is based on maximizing a quality function called modularity~\cite{newman2004finding}. Another well-accepted framework to study communities is to consider the stochastic block model (SBM)~\cite{abbe2017community}, which is a random graph model producing graphs comprising communities. 

Since opinion evolution depends on the underlying network topology, it is reasonable to believe that community structure of a network may greatly influence opinion dynamics over that network. One such intuition is that agents in the same community should have similar opinions. This phenomenon has been observed in simulations for many models such as the DW model~\cite{gargiulo2010opinion,fennell2021generalized}, the Taylor model (a continuous-time version of the FJ model)~\cite{baumann2020laplacian}, the Sznajd model~\cite{si2009opinion}, and so on. The paper~\cite{como2016local} shows that, for the DeGroot model over a special weighted graph, the steady states of agents in the same community concentrate around the state of some stubborn agent.

\subsection{Contribution}

In this paper, we study transient behavior of the gossip model over a weighted graph with two communities. It is assumed that edges within communities have identical weights different from the weights between communities. It is shown that, when the edge weights within communities are larger and the influence of stubborn agents is small, agents have expected states with identical sign corresponding to their community labels over a transient time interval (Theorem~\ref{thm_expectation}). Over this time interval, the number of updates per agent is of the same order as the ratio of the edge weights within communities and the weights between communities. Under the same condition, we further verify that the difference between the expected state of an agent and the initial average opinion of its community is bounded by a quantity depending on time and the eigenvalues of the weighted adjacency matrix. Hence, over a transient time interval similar to Theorem~\ref{thm_expectation}, agents have expected states close to the initial average opinion of their communities (Theorem~\ref{thm_state}~(i)). In contrast, if the edge weights between communities are larger, then  the expected states of all agents concentrate around the initial average opinion of the entire network (Theorem~\ref{thm_state}~(ii)).


These results indicate that the gossip model completely changes its behavior at the early stage, under different levels of intra-community and inter-community interaction strength. 
It is known that 
stationary agent states depend on stubborn-agent states~\cite{acemouglu2013opinion}. Since the obtained results indicate that the transient clusters only depend on the initial states and the edge weights between agents, 
transient behavior of the model can be different from asymptotic behavior. These observations deepen our understanding of the opinion formation process. The results also establish a correspondence between community structure and transient behavior of the model, and validate the intuition that transient clusters can appear over a network with community structure.

The results explain how community structure influences the emergence and the duration of transient clusters in opinion dynamics, and thus provide insights into predicting and distinguishing such transient phenomena in practice~\cite{banisch2012agent,banisch2010empirical}. The correspondence between community structure and transient behavior indicates that it is possible to develop community-detection methods for dynamic systems based on state observations~\cite{schaub2020blind,xing2021detecting}. Finally, exploiting properties of transient clusters can help improve model reduction tools at the initial phase of dynamics over large-scale networks with community structure~\cite{cheng2018model,niazi2020state}.


\subsection{Outline}
In Section~\ref{sec_problem} we introduce the considered problem. Section~\ref{sec_results} provides main results of the paper, and Section~\ref{sec_proofs} gives proofs of the results. Numerical experiments are presented in Section~\ref{sec_simulation}. 

\noindent\textbf{Notation.}
Denote the $n$-dimensional Euclidean space by $\mathbb{R}^n$, the set of $n\times m$ real matrices by $\mathbb{R}^{n\times m}$, the set of nonnegative integers by $\mathbb{N}$, and $\mathbb{N}^+ = \mathbb{N}\setminus\{0\}$. $\log x$, $x\in\RR$, is the natural logarithm. Let $\mathbf{1}_n$ be the all-one vector with dimension $n$, $\textbf{e}_1$, $\dots$, $\textbf{e}_n$ be the canonical basis of $\RR^n$, $I_n$ be the $n\times n$ identity matrix (we omit $n$ if there is no confusion). Denote the Euclidean norm of a vector by $\|\cdot\|$. 
For a vector $x\in \mathbb{R}^n$, denote its $i$-th entry by $x_i$, and for a matrix $A \in \mathbb{R}^{n\times n}$, denote its $(i,j)$-th entry by $a_{ij}$ or $[A]_{ij}$. 
The cardinality of a set $\mtcs$ is denoted by $|\mtcs|$. The function $\mathbb{I}_{[\textup{property}]}$ is the indicator function equal to one if the property in the bracket holds, and equal to zero otherwise.  
Denote the expectation of a random vector $X$ by $\mathbb{E}\{X\}$.  For two sequences of real numbers, $f(n)$ and $g(n) > 0$, $n\in\NN$, we write $f(n) = O_n(g(n))$ if $|f(n)| \le C g(n)$ for all $n\in \NN$, $f(n) = o_n(g(n))$ if $|f(n)|/g(n) \to 0$, and $f(n) \sim g(n)$ if $f(n)/g(n) \to 1$ as $n \to \infty$. Further assuming that $f(n)$, $n\in\NN$, are nonnegative, we say $f(n) = \omega_n(g(n))$ if $g(n) = o_n(f(n))$, $f(n) = \Omega_n(g(n))$ if there is $C>0$ such that $f(n) \ge C g(n)$, $n \in \NN$, and $f(n) = \Theta_n(g(n))$ if both $f(n) = O_n(g(n))$ and $f(n) = \Omega_n(g(n))$ hold. The subscript $n$ is omitted if there is no confusion.

\begin{figure*}[t]
	\centering
	\includegraphics[scale=0.4]{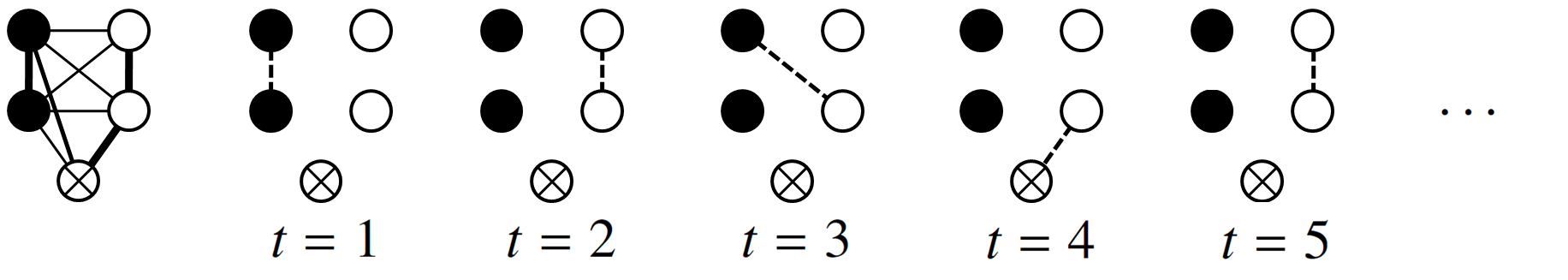}
	\caption{\label{fig_illus_asmp} Illustration of Assumption~\ref{asmp_1}~(i). The graph on the left demonstrates the underlying network with two communities (dots and circles) and one stubborn agent  (the circle with a cross). Solid lines represent weighted edges. The weights are indicated by line thickness. In the network, edge weights within communities are larger than between communities ($l_s^{(r)} > l_d^{(r)}$). The rest of the graphs in the figure show random interactions between agents, represented by dashed lines, at each time step. Agents interact more often if they have an edge with a larger weight.}	
\end{figure*}

\section{Problem Formulation}\label{sec_problem}

The gossip model with stubborn agents is a random process evolving over a graph $\mathcal{G} = (\mathcal{V}, \mathcal{E},A)$, where $\mtcv$ is the node set with $|\mathcal{V}| = n \ge 2$, $\mtce$ is the edge set, and $A$ is the weighted adjacency matrix.
The graph $\mathcal{G}$ has no self-loops. In addition, $\mathcal{V}$ contains two types of agents, regular and stubborn, denoted by $\mathcal{V}_r$ and $\mathcal{V}_s$, respectively ($\mathcal{V} = \mathcal{V}_r \cup \mathcal{V}_s$ and $\mathcal{V}_r \cap \mathcal{V}_s = \emptyset$). Each agent~$i$ in the graph possesses a state $Z_i(t) \in \mathbb{R}$, $t \in \mathbb{N}$. Stacking all states, we denote the state vector at time $t$ by $Z(t) \in \mathbb{R}^n$.  

The random interaction of the gossip model is captured by an interaction probability matrix $W = [w_{ij}] \in \mathbb{R}^{n\times n}$ satisfying that $w_{ij} = w_{ji} = a_{ij}/\alpha$, where $\alpha = \sum_{i=1}^n\sum_{j=i+1}^n a_{ij}$ is the sum of all edge weights of $\mtcg$. Hence $\mathbf{1}^T W \mathbf{1}/2 = 1$.

 At time $t$, edge $\{i,j\}$ is selected with probability $w_{ij}$ independently of previous updates, and agents update as follows, 
\begin{align}\label{eq_update_rule1}
    Z_k(t+1) = \begin{cases}
            \frac12 (Z_i(t) + Z_j(t)), & \text{if } k \in \mathcal{V}_r \cap \{i,j\},\\
            Z_k(t), & \text{otherwise}.
            \end{cases}
\end{align}
The averaging weight is assumed to be $1/2$, but general weights can be considered. 
For $1 \le i < j \le n$, define
\begin{equation*}
    V^{ij} = \begin{cases}
    I - \frac12 (\textbf{e}_i - \textbf{e}_j)(\textbf{e}_i - \textbf{e}_j)^T, & \text{if } i, j \in \mathcal{V}_r,\\
    I - \frac12 \textbf{e}_i(\textbf{e}_i - \textbf{e}_j)^T, & \text{if } i \in \mathcal{V}_r,  j \in \mathcal{V}_s,\\
    I - \frac12 \textbf{e}_j(\textbf{e}_j - \textbf{e}_i)^T, & \text{if } i \in \mathcal{V}_s,  j \in \mathcal{V}_r,\\
    I, & \text{if } i, j \in \mathcal{V}_s,
    \end{cases}
\end{equation*}
and a sequence of independent and identically distributed (i.i.d.) $n$-dimensional random matrices $\{V(t),$ $t\in \mathbb{N}\}$ such that $ \mathbb{P}\{V(t) = V^{ij}\} = w_{ij}$, $1 \le i < j \le n.$
The compact form of update rule~\eqref{eq_update_rule1} is
\begin{align}\label{eq_update_compact}
    Z(t+1) = V(t) Z(t).
\end{align}
Since stubborn agents never change their states during the process, we rewrite~\eqref{eq_update_compact} and get the following compact form of the gossip model:
\begin{align}\label{eq_update_compact_regular}
    X(t+1) = Q(t) X(t) + R(t) \mathbf{z}^s,
\end{align}
where $X(t)$ and $\mathbf{z}^s$ denote the state vectors obtained by stacking the states of regular and stubborn agents, respectively, and $[Q(t) ~ R(t)]$ is the matrix obtained by stacking rows of $V(t)$ corresponding to regular agents. 

In this paper, we assume that the regular agents form two disjoint communities $\mtcv_{r1}$ and $\mtcv_{r2}$ with equal size (hence $\mtcv_r = \mtcv_{r1} \cup \mtcv_{r2}$), and denote $\mtcc_i = k$ if $i \in \mtcv_{rk}$, $k=1,2$. For simplicity, number the agents as follows: $\mtcv_{r1} = \{1,\dots, r_0n/2\}$, $\mtcv_{r2} = \{1 + r_0n/2,\dots, r_0n\}$, and $\mtcv_s = \{1+r_0n, \dots, n\}$, with $r_0 \in (0,1)$ such that $r_0n$ is an even integer. The portion of stubborn agents is denoted by $s_0 := 1 - r_0$.

We introduce the following assumptions for the graph $\mtcg$ and the initial vector of the gossip model~\eqref{eq_update_compact_regular}.

\begin{asmp}\label{asmp_1}~
	\begin{enumerate}[(i)]
		\item There exist constants $l_s^{(r)}, l_d^{(r)}, l_{ij}^{(s)} \in (0,1)$, $1\le i \le r_0n$, $1\le j\le s_0n$, such that $a_{ij} = l_s^{(r)}$ for $i,j\in \mtcv_r$ with $i\not=j$ and $\mtcc_i = \mtcc_j$, $a_{ij} = l_d^{(r)}$ for $i,j\in \mtcv_r$ with $\mtcc_i \not= \mtcc_j$, and $a_{i, r_0n+j} = a_{r_0n+j, i} = l_{ij}^{(s)}$ for $i \in \mtcv_r$ and $1\le j \le s_0n$. Other entries of $A$ are zero.
		\item There exists a positive constant $l^{(s)}$ such that $\sum_{1\le j \le s_0n} l_{ij}^{(s)} = l^{(s)}$ for all $i \in \mtcv_r$.
		\item The initial vector $X(0)$ and the stubborn-agent states $\bfz^s$ are deterministic, and satisfy that $|X_i(0)| \le c_x$ and $|\bfz^s_j| \le c_x$, for all $1\le i \le r_0n$ and $1\le j\le s_0n$, and some constant $c_x>0$.\hspace*{\fill}~\QED\par\endtrivlist\unskip
	\end{enumerate}
\end{asmp}

In what follows, we give some remarks on the preceding assumptions.

\begin{rmk}
	 Assumption~\ref{asmp_1}~(i), illustrated in Fig.~\ref{fig_illus_asmp}, implies that the graph on regular agents is a weighted complete graph, where edges between agents in the same community have the same weight $l_s^{(r)}$ and edges between communities have weight $l_d^{(r)}$.  The assumption of the weighted complete graph is representative because this graph can be considered as the expectation of an SBM with two communities (see Example~1 of~\cite{xing2021detecting} for details). $l^{(s)}$ in~(ii) is the sum of edge weights between a regular agent and stubborn agents, and thus represents the total influence of stubborn agents on this regular agent. We assume that this sum is the same for all regular agents. This assumption is introduced for analysis simplicity, but can be extended if the upper and lower bounds of the weight sums are available. In~(iii) it is assumed that the initial values of all agents are bounded, and thus the process is bounded as well.\hspace*{\fill}~\QED\par\endtrivlist\unskip
\end{rmk}

In this paper, we investigate how community structure (the collection of community labels $\mtcc_i$) influences the expected agent states during the early stage of the process. The problem considered is as follows.

\textbf{Problem.} Given the initial vector $X(0)$, the stubborn-agent states $\bfz^s$, the community structure $\mtcc$, and the weighted adjacency matrix $A$, for all agent $i \in \mtcv_r$ and time $t$ in some transient interval, characterize the sign and the value of $\EE\{X_i(t)\}$.\hspace*{\fill}~\QED\par\endtrivlist\unskip

The sign of  $\EE\{X_i(t)\}$ is studied in Theorem~\ref{thm_expectation}, and the value of $\EE\{X_i(t)\}$ is investigated in Theorem~\ref{thm_state} in the next section.

\section{Main Results}\label{sec_results}
In this section, we provide the main results for transient behavior of the gossip model~\eqref{eq_update_compact_regular}. The proofs of the results are given in Section~\ref{sec_proofs}.

Note that, from the definitions of $Q(t)$ and $R(t)$ in~\eqref{eq_update_compact_regular}, it follows that
\begin{align}\nonumber
	\bar{Q} &\Let \EE\{Q(t)\} \\\label{eq_barQ}
	&= I - \frac{1}{2\alpha}
	\begin{bmatrix}
	d_1 & -a_{12} & \cdots & -a_{1,r_0n}\\
	-a_{21} & \ddots & & \vdots\\
	\vdots &  & & \vdots \\
	-a_{r_0n,1} & \cdots & -a_{r_0n,r_0n-1} & d_{r_0n}
	\end{bmatrix}, \\\label{eq_barR}
	\bar{R} &\Let \EE\{R(t)\} = \frac{1}{2\alpha} \tilde{M} \Let \frac{1}{2\alpha} 
	\begin{bmatrix}
		a_{1,r_0n+1} & \cdots & a_{1, n}\\
		\vdots & & \vdots\\
		a_{r_0n, r_0n+1} & \cdots & a_{r_0n,n}
	\end{bmatrix},
\end{align}
where $d_i = \sum_{j\in \mtcv} a_{ij}$, $i \in \mtcv_r$. Denote $\lambda_1 \Let  l^{(s)}/(2\alpha)$, $\lambda_2 \Let (l_d^{(r)} r_0n + l^{(s)})/(2\alpha)$, and $\lambda_3 \Let [(l_s^{(r)} + l_d^{(r)}) r_0n/2 + l^{(s)}]/(2\alpha)$. Since $\bar{Q} \in \RR^{r_0n\times r_0n}$ is symmetric, it has an eigenvalue $1-\lambda_1$ with unit eigenvector $\eta \Let \bfl_{r_0n}/\sqrt{r_0n}$, an eigenvalue $1 - \lambda_2$ with unit eigenvector $\xi := [\bfl_{r_0n/2}^T~-\bfl_{r_0n/2}^T]/\sqrt{r_0n}$, and an eigenvalue $1 - \lambda_3$ with unit eigenvectors $w^{(i)}$, $3\le i \le r_0n$. Moreover, $\eta$, $\xi$, $w^{(3)}$, $\dots$, $w^{(r_0n)}$ form an orthonormal basis of $\RR^{r_0n}$. Let us introduce a little more notation: $c_{\eta,x} \Let \eta^T X(0)$, $c_{\xi, x} \Let \xi^T X(0)$, $\zeta_1 \Let \eta^T \tilde{M} \bfz^s / l^{(s)}$, and $\zeta_2 \Let \xi^T \tilde{M} \bfz^s / l^{(s)}$. 

Now we are ready to state the first result, which concerns how the sign of $\EE\{X(t)\}$ is influenced by  community labels at the beginning of the process.

\begin{thm}\label{thm_expectation}
	Suppose that Assumption~\ref{asmp_1} holds and $l_s^{(r)} > l_d^{(r)}$. Let 
	\begin{align*}
		\underline{t} &:= \frac{\log (15 c_x \sqrt{r_0n} / |c_{\xi,x}|)}{\log [(1 - \lambda_2)/(1 - \lambda_3)]},\\
		\overline{t} &:= \min \bigg\{ \frac{\log (|c_{\xi,x}| / 5 |c_{\eta,x}|)}{\log [(1 - \lambda_1)/(1 - \lambda_2)]}, \frac{\log (|c_{\xi,x}|/5|\zeta_1|)}{\log [1/(1 - \lambda_2)]}, \\
			&\qquad \frac{ \log \{[(l_d^{(r)} r_0 n + l^{(s)}) |c_{\xi,x}| / 5 l^{(s)} |\zeta_2|] +1 \} }{\log [1/(1-\lambda_2)]},\\
			&\qquad \frac{\log \{[\frac12(l_s^{(r)} + l_d^{(r)}) r_0n + l^{(s)}] |c_{\xi,x}| / 15 l^{(s)} c_x \sqrt{r_0n}\}}{\log [1/ (1 - \lambda_2)]} \bigg\}.
	\end{align*}
	If $c_{\xi,x} \not=0$, $(\underline{t}, \overline{t})$ is nonempty, and $n$ is large enough, then $\sgn (\EE\{X_i(t)\}) = \sgn (c_{\xi,x} \xi_i)$ holds for all $t \in (\underline{t}, \overline{t})$ and $i \in \mtcv_r$.\hspace*{\fill}~\QED\par\endtrivlist\unskip
\end{thm}

\begin{rmk}
	The preceding result shows that, if the time interval $t \in (\underline{t}, \overline{t})$ is nonempty, then within this interval the sign of $\EE\{X_i(t)\}$ is the same as that of $c_{\xi,x}\xi_i$, for all $i\in \mtcv_r$. Note that the sign of $\xi_i$ is positive if $i \in \mtcv_{r1}$, and is negative if $i \in \mtcv_{r2}$. Thus Theorem~\ref{thm_expectation} implies that the expectation of $X_i(t)$ for all $i$ in the same community is identical, during a short period after the process begins. The result establishes a preliminary relationship between community structure and  transient behavior of the gossip model. To ensure that $t \in (\underline{t}, \overline{t})$ is nonempty, $c_{\xi,x}$ needs be large enough (meaning that the initial average opinion difference between different communities is large enough), and $c_{\eta,x}$ needs be small enough (intuitively, the initial average opinion is small). In addition, the influence of stubborn agents should also be small (that is, $|\zeta_1|$ and $|\zeta_2|$ are small). It should be noted that the result is still valid when there are no stubborn agents  (i.e., $\bfz^s$ or $\tilde{M}$ is zero), and the upper bound $\overline{t}$ of the time interval becomes $  [\log (|c_{\xi,x}| / 5 |c_{\eta,x}|)]/\{\log [(1 - \lambda_1)/(1 - \lambda_2)]\}$. \hspace*{\fill}~\QED\par\endtrivlist\unskip
\end{rmk}

In the following example, we show that the interval $(\underline{t}, \overline{t})$ can be nonempty.

\begin{exmp}\label{exmp_expectation}
	To examine the possibility that the interval $(\underline{t}, \overline{t})$ is nonempty, in this example, we assume that $X(0)$ is generated from some random vector such that $X_i(0) \in [0,c_x]$, $i\in \mtcv_{r1}$, are i.i.d. random variables with mean $c_0 > 0$, and $\{X_i(0), i \in \mtcv_{r2}\}$, are i.i.d. random variables that have distribution $-X_1(0)$ and are independent of $\{X_i(0), i \in \mtcv_{r1}\}$. Hence, from the Hoeffding inequality~\cite{vershynin2018high}, it follows that $c_{\xi,x} = \Theta(\sqrt{r_0n})$ and $c_{\eta,x} = O(1)$ with  positive probability. Now assume that $l_s^{(r)} = \omega(l_d^{(r)})$ and $l_s^{(r)}n = \omega(l^{(s)})$ as $n \to \infty$. Thus, $\alpha = r_0n[(l_s^{(r)} + l_d^{(r)}) r_0 n + 4l^{(s)} -2l_s^{(r)}]/4 = \Theta( n^2 l_s^{(r)})$, $\log [1/(1-\lambda_2)] = \Theta (l_d^{(r)}/n l_s^{(r)})$, $\log [(1-\lambda_1)/(1-\lambda_2)] = \Theta (l_d^{(r)}/n l_s^{(r)})$, and $\log [(1-\lambda_2)/(1-\lambda_3)] = \Theta (1/n)$. Thus, $\underline{t} = \Theta(n)$. Furthermore, from Assumption~\ref{asmp_1}~(ii), we know that $|\zeta_1| \le c_x\sqrt{r_0n}$ and $|\zeta_2| \le c_x\sqrt{r_0n}$. Thus, $\overline{t} = \Omega(n  l_s^{(r)}/l_d^{(r)})$. Therefore, within the time interval $(c_1 n, c_2n  l_s^{(r)}/l_d^{(r)})$ for some $c_1,c_2>0$, the result of Theorem~\ref{thm_expectation} holds with positive probability (i.e., the sign of $\EE\{X_i(t)\}$ corresponds to $\mtcc_i$, for all $i\in \mtcv_r$).
		
	In particular, consider $l_s^{(r)} = (\log^{\beta_1} n)/n$, $l_d^{(r)} = (\log^{\beta_2} n)/n$, and $l^{(s)} = \log^{\beta_3} n$ with $\beta_1 > \beta_2 \ge \beta_3 \ge 1$. These parameter values correspond to dense SBMs, which generate connected graphs with high probability~\cite{abbe2017community}. Under such conditions, the time interval is $(c_1 n, c_2n  \log^{\beta_1-\beta_2} n )$. Since there are at most two agents updating at each time step, over such a time interval, regular agents update for approximately $\log^{\beta_1-\beta_2} n$ times.
	\hspace*{\fill}~\QED\par\endtrivlist\unskip
\end{exmp}

\begin{rmk}
	The preceding example shows that, if the initial opinion difference between agents in different communities is large
	, and the edges between communities and between regular and stubborn agents have relatively small weights, then agents in the same community would keep an identical sign on average, during the early stage of the gossip process. Theorem~\ref{thm_expectation} and Example~\ref{exmp_expectation} establish this correspondence and show the duration of such behavior. 
	\hspace*{\fill}~\QED\par\endtrivlist\unskip
\end{rmk}

In Theorem~\ref{thm_expectation}, we showed a correspondence between community labels and expected agent states. The next theorem provides stronger results, which indicate that the expected states in the same community are close to each other over a transient time interval.


\begin{thm}\label{thm_state}
	Suppose that Assumption~\ref{asmp_1} holds. \\
	(i) If $l_s^{(r)} > l_d^{(r)}$, then for all $i\in \mtcv_r$ and $t\in\NN$, it holds that
	\begin{align*}
		&\bigg|\EE\{X_i(t)\} - \frac{2}{r_0n} \sum_{j \in \mtcv_{r\mtcc_i}} X_j(0) \bigg|\\
		& \le [(4\lambda_1 + \lambda_2)t + (1 - \lambda_3)^t]c_x.
	\end{align*}
	Further suppose that, for $\underline{t}_1$ and $\overline{t}_1$ such that $\underline{t}_1 =\underline{t}_1(n) = \omega_n(1)/\lambda_3$ and $\overline{t}_1 = \overline{t}_1(n) = o_n(1/\lambda_2)$, $\underline{t}_1 = o_n(\overline{t}_1)$ holds. Then, for all $i\in \mtcv_r$ and $t \in (\underline{t}_1,\overline{t}_1)$, it holds that
	\begin{align}\label{eq_thm_state_local}
		\bigg|\EE\{X_i(t)\} - \frac{2}{r_0n} \sum_{j \in \mtcv_{r\mtcc_i}} X_j(0) \bigg| = o_n(1).
	\end{align}
	(ii) If $l_s^{(r)} \le l_d^{(r)}$, then for all $i\in \mtcv_r$ and $t\in\NN$, it holds that
	\begin{align*}
		&\bigg|\EE\{X_i(t)\} - \frac{1}{r_0n} \sum_{j \in \mtcv_{r}} X_j(0) \bigg|\\
		& \le [4\lambda_1t + 2(1 - \lambda_3)^t]c_x.
	\end{align*}
	Further suppose that, for $\underline{t}_2$ and $\overline{t}_2$ such that $\underline{t}_2 =\underline{t}_2(n) = \omega_n(1)/\lambda_3$ and $\overline{t}_2 = \overline{t}_2(n) = o_n(1/\lambda_1)$, $\underline{t}_2 = o_n(\overline{t}_2)$ holds. Then, for all $i\in \mtcv_r$ and $t \in (\underline{t}_2,\overline{t}_2)$, it holds that
	\begin{align}\label{eq_thm_state_global}
		\bigg|\EE\{X_i(t)\} - \frac{1}{r_0n} \sum_{j \in \mtcv_{r}} X_j(0) \bigg| = o_n(1). 
	\end{align}
\end{thm}

\begin{rmk}\label{rmk_thm_state}
	The first part of the theorem states that, if connections within communities are stronger than between communities ($l_s^{(r)} > l_d^{(r)}$) and the influence of stubborn agents is relatively small ($\underline{t}_1 = o_n(\overline{t}_1)$), the agents in the same community are close to a local consensus and have expected states close to their initial average opinion, over a transient time interval, depending on the network size $n$. From Example~\ref{exmp_expectation}, each agent updates for $l_s^{(r)}/l_d^{(r)}$ times on average. In contrast, the second part of the theorem implies that the agents are close to a global consensus over a transient time interval, when connections between communities are stronger. In this case, each agent updates for $l_d^{(r)}/l^{(s)}$ times on average. It should be noted that the assumption $l_s^{(r)} \le l_d^{(r)}$ can be relaxed to $l_s^{(r)} = O(l_d^{(r)})$, which implies that the system is close to a global consensus even when edge weights within communities are slightly larger than weights between communities. These two results indicate completely different transient behavior for the model with different community structure. \hspace*{\fill}~\QED\par\endtrivlist\unskip 
\end{rmk}
\begin{rmk}	
	It is known that, for the gossip model, the expected stationary states depend on the positions of stubborn agents~\cite{acemouglu2013opinion}, and the system reaches a consensus if there are no stubborn agents~\cite{boyd2006randomized,fagnani2008randomized}. Theorem~\ref{thm_state} shows that the agents can form transient clusters, which may be different from their asymptotic behavior, if the influence of stubborn agents is relatively small. The positions of the transient clusters depend only on initial states and link strength between regular agents. This result provides conditions for the existence and the duration of transient clusters emerging over a network with community structure. \hspace*{\fill}~\QED\par\endtrivlist\unskip 
\end{rmk}

	Theorem~\ref{thm_state} has several potential applications. Based on the duration of a transient cluster given by the theorem, we may predict transient behavior of a process~\cite{banisch2012agent,banisch2010empirical}. In addition, the first part of the theorem implies that agents in the same community have similar states if they have stronger links between each other. This observation can provide insights into designing community detection methods based on state observations~\cite{schaub2020blind,xing2021detecting}, and into improving model reduction tools in the initial phase for dynamics over large-scale networks with community structure~\cite{cheng2018model,niazi2020state}.

We conclude this section by providing an example for Theorem~\ref{thm_state}.

\begin{exmp}\label{exmp_state}
	Following the discussion in Example~\ref{exmp_expectation}, let $l_s^{(r)} = (\log^{\beta_1} n)/n$, $l_d^{(r)} = (\log^{\beta_2} n)/n$, and $l^{(s)} = \log^{\beta_3} n$ with $\beta_1 > \beta_2 \ge \beta_3 \ge 1$, then $1/\lambda_1 = \Theta(n(\log n)^{\beta_1-\beta_3})$, $1/\lambda_2 = \Theta(n(\log n)^{\beta_1-\beta_2})$, and $1/\lambda_3 = \Theta(n)$. We can select $f(n)$ that grows slowly to infinity, say $f(n) = \log\log n$, and then $\underline{t}_1 \approx c_1n$ for small $n$ and $c_1>0$. In addition, set $\overline{t}_1 = n (\log n)^{(\beta_1-\beta_2)/2}$. So~\eqref{eq_thm_state_local} holds for $t \in (c_1n, n (\log n)^{(\beta_1-\beta_2)/2})$. In contrast, if $\beta_3 \le \beta_1 < \beta_2$, then $l_s^{(r)} < l_d^{(r)}$, and~\eqref{eq_thm_state_global} holds for $t \in (c_1n, n (\log n)^{(\beta_2-\beta_3)/2})$.
\end{exmp}

\begin{rmk}
	In the case where $\beta_1 > \beta_2 \ge \beta_3 \ge~1$ in Example~\ref{exmp_state}, the transient interval $(n, n (\log n)^{(\beta_1-\beta_2)/2})$, in which~\eqref{eq_thm_state_local} holds, shrinks when $\beta_1$ decreases. When $\beta_1$ becomes smaller than $\beta_2$, the transient interval $(n, n (\log n)^{(\beta_2-\beta_3)/2})$, in which~\eqref{eq_thm_state_global} holds, expands as $\beta_2$ increases. 
	\hspace*{\fill}~\QED\par\endtrivlist\unskip
\end{rmk}

\section{Proofs}\label{sec_proofs}
In this section, we provide proofs of the results given in Section~\ref{sec_results}. Denote $x(t) \Let \EE\{X(t)\}$ for simplicity.

\subsection{Proof of Theorem~\ref{thm_expectation}} 
Note that, from the discussion after~\eqref{eq_barR}, it follows that $\bar{Q}$ can be written as
\begin{align*}
	(1 - \lambda_1) \eta \eta^T +  ( 1 - \lambda_2) \xi \xi^T + ( 1 - \lambda_3) \sum_{i = 3}^{r_0n} w^{(i)} (w^{(i)})^T.
\end{align*}
Hence, we have that
\begin{align}\nonumber
	&x(t) =
	 \Big\{ ( 1 -  \lambda_1)^t (\eta^Tx(0)) + \frac{1}{ \lambda_1}  [1 - ( 1 -  \lambda_1)^t  ] (\eta^T\bar{R} \bfz^s) \Big\} \eta  \\\nonumber
	& + \Big\{ ( 1 -  \lambda_2)^t (\xi^Tx(0)) + \frac{1}{\lambda_2}  [1 - ( 1 - \lambda_2)^t ] (\xi^T\bar{R} \bfz^s) \Big\} \xi \\\nonumber
	& + \sum_{i = 3}^{r_0n} \Big\{ ( 1 - \lambda_3)^t ((w^{(i)})^Tx(0))  + \frac{1}{\lambda_3} [ 1-  ( 1 - \lambda_3)^t  ]\\\label{eq_expression_Ext}
	&
	 ((w^{(i)})^T\bar{R} \bfz^s) \Big\} w^{(i)}.
\end{align}
To verify the conclusion, it suffices to validate that the entries of the term $( 1 -  \lambda_2)^t (\xi^Tx(0)) \xi$ are larger than those of the other terms in~\eqref{eq_expression_Ext} in absolute value, under the conditions of the theorem.
\hspace*{\fill}~\QED\par\endtrivlist\unskip

\subsection{Proof of Theorem~\ref{thm_state}} 
To prove~(i) of Theorem~\ref{thm_state}, first note that $2 (\sum_{j \in \mtcv_{r\mtcc_i}} X_j(0))/(r_0n) = (\eta^T X(0)) \eta_i + (\xi^T X(0)) \xi_i$ for all $i\in \mtcv_r$, so it follows from~\eqref{eq_expression_Ext} that
\begin{align}\nonumber
	&\bigg|x_i(t) - \frac{2}{r_0n} \sum_{j \in \mtcv_{r\mtcc_i}} X_j(0) \bigg| \\\nonumber
	&= |x_i(t) - [(\eta^T X(0)) \eta_i + (\xi^T X(0)) \xi_i ]| \\\nonumber
	&\le 
	[1 - (1 - \lambda_1)^t] |(\eta^T X(0)) \eta_i| + [1 - (1 - \lambda_1)^t] |\zeta_1 \eta_i| \\\nonumber
	&\quad + [1 - (1 - \lambda_2)^t] |(\xi^T X(0)) \xi_i| +  \frac{\lambda_1}{\lambda_2}[1 - (1 - \lambda_2)^t] |\zeta_2 \xi_i| \\\nonumber
	&\quad +(1 - \lambda_3)^t \bigg| \bigg(\sum_{j = 3}^{r_0n} w^{(j)}  ((w^{(j)})^Tx(0)) \bigg)_i\bigg| \\\nonumber
	&\quad + \frac{\lambda_1}{\lambda_3}  [ 1-  ( 1 - \lambda_3)^t  ] \bigg| \bigg( \sum_{j = 3}^{r_0n} w^{(j)}  ((w^{(j)})^T \bar{R} \bfz^s/\lambda_1) \bigg)_i\bigg| \\\nonumber
	&\le
	c_x \bigg\{ 2[1 - (1 - \lambda_1)^t]   + [1 - (1 - \lambda_2)^t]  \\\nonumber
	&\quad +  \frac{\lambda_1}{\lambda_2}[1 - (1 - \lambda_2)^t]  + (1 - \lambda_3)^t   + \frac{\lambda_1}{\lambda_3}  [ 1-  ( 1 - \lambda_3)^t  ]  \bigg\}\\\label{eq_thm_state_i}
	&\le [(4\lambda_1  + \lambda_2) t + (1-\lambda_3)^t ]c_x,
\end{align}
where $\zeta_1$ and $\zeta_2$ are given before Theorem~\ref{thm_expectation}, the penultimate inequality follows from Assumption~\ref{asmp_1}~(iii), and the last inequality is obtained from the Bernoulli inequality
\begin{align*}
	(1+x)^r \ge 1 + rx, \forall r \ge 1, x \ge -1.
\end{align*}
When $t \in (\underline{t}_1,\overline{t}_1)$, $(4\lambda_1  + \lambda_2) t \le 5 \lambda_2 t < 5\lambda_2 \overline{t}_1 = o_n(1)$, and 
\begin{align*}
	(1-\lambda_3)^t = (1-\lambda_3)^{-\frac{1}{\lambda_3} (-\lambda_3 t)} \sim e^{-\lambda_3 t} \le e^{-\lambda_3 \underline{t}_1} = o_n(1).
\end{align*}
Thus, $\eqref{eq_thm_state_i} = c_x o_n(1) = o_n(1)$.

To verify (ii) of Theorem~\ref{thm_state}, it suffices to note that
\begin{align}\nonumber
	&\bigg|x_i(t) - \frac{1}{r_0n} \sum_{j \in \mtcv_{r}} X_j(0) \bigg| \\\nonumber
	&= |x_i(t) - (\eta^T X(0)) \eta_i | \\\nonumber
	&\le 
	[1 - (1 - \lambda_1)^t] |(\eta^T X(0)) \eta_i| + [1 - (1 - \lambda_1)^t] |\zeta_1\eta_i| \\\nonumber
	&\quad +  (1 - \lambda_2)^t|(\xi^T X(0)) \xi_i| +  \frac{\lambda_1}{\lambda_2}[1 - (1 - \lambda_2)^t] |\zeta_2\xi_i| \\\nonumber
	&\quad +(1 - \lambda_3)^t \bigg| \bigg(\sum_{j = 3}^{r_0n} w^{(j)}  ((w^{(j)})^Tx(0)) \bigg)_i\bigg| \\\nonumber
	&\quad + \frac{\lambda_1}{\lambda_3}  [ 1-  ( 1 - \lambda_3)^t  ] \bigg| \bigg( \sum_{j = 3}^{r_0n} w^{(j)}  ((w^{(j)})^T \bar{R} \bfz^s/\lambda_1) \bigg)_i\bigg| \\\nonumber
	&\le
	c_x \bigg\{ 2[1 - (1 - \lambda_1)^t]   + (1 - \lambda_2)^t +  \frac{\lambda_1}{\lambda_2}[1 - (1 - \lambda_2)^t] \\\nonumber
	&\quad  + (1 - \lambda_3)^t   + \frac{\lambda_1}{\lambda_3}  [ 1-  ( 1 - \lambda_3)^t  ]  \bigg\}\\\label{eq_thm_state_ii}
	&\le [4\lambda_1 t + 2(1-\lambda_3)^t ]c_x,
\end{align}
where the last inequality holds because $l_s^{(r)} \le l_d^{(r)}$ implies that $1 - \lambda_2 \le 1 - \lambda_3$. \hspace*{\fill}~\QED\par\endtrivlist\unskip

\addtolength{\textheight}{-7.5cm}   

\section{Numerical Simulation}\label{sec_simulation}
In this section, we conduct numerical experiments to validate the theoretical results obtained in Section~\ref{sec_results}. We first show that the expected agent states in the same community have an identical sign at the early stage of the process, and they are close to the initial average opinion of that community during the transient phase, when connections within communities are stronger. Then we demonstrate that, in contrast, all agents have similar expected states, when connections between communities are stronger. Finally, we show that similar phenomenon exists for agent states.

\begin{figure}[t]
	\centering
	\includegraphics[scale=0.38]{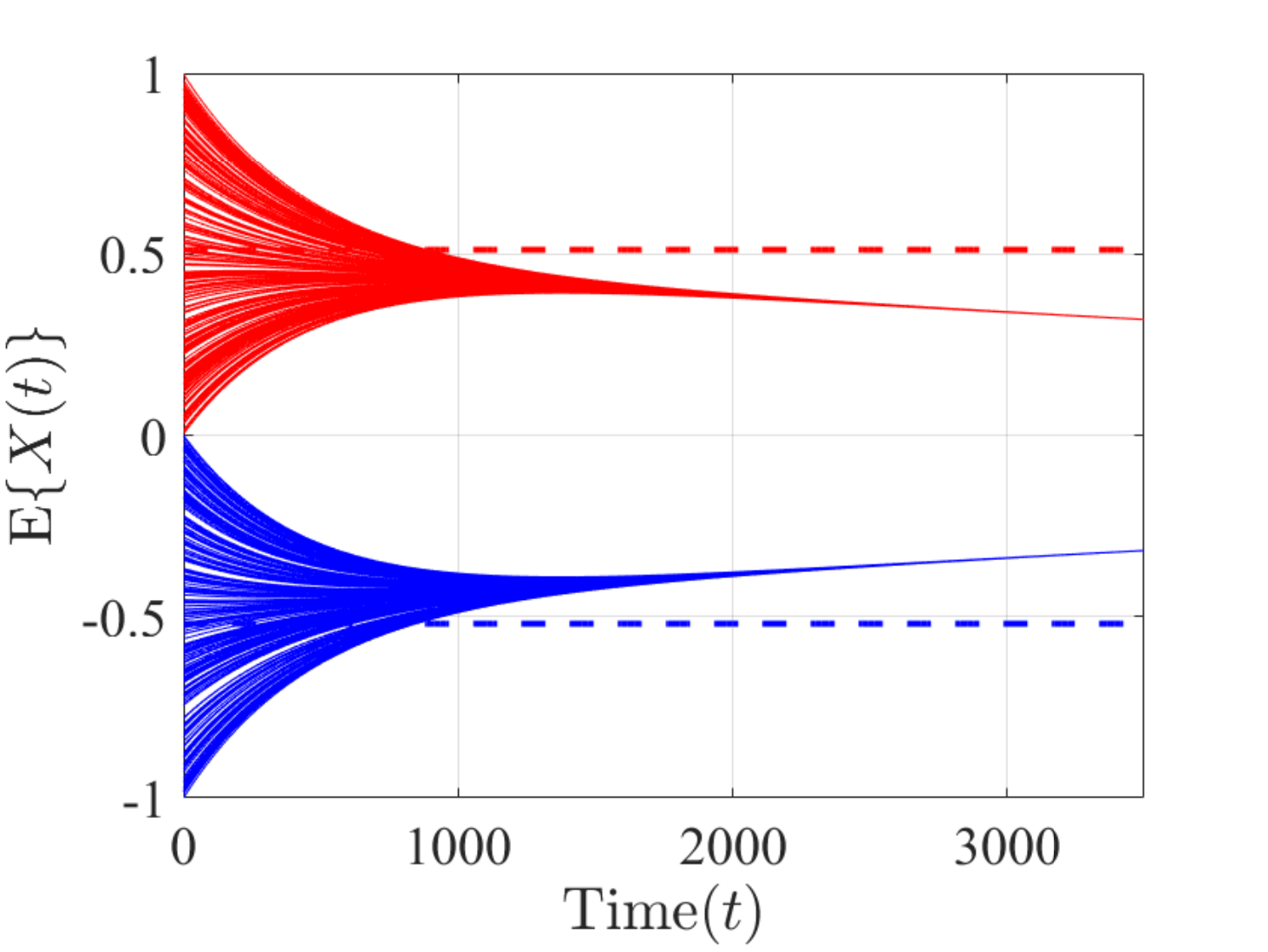}
	\caption{\label{fig_exp_local} Behavior of expected states when $l_s^{(r)} > l_d^{(r)}$. The expected states of agents in the same community are close and have the same sign over the time interval $(n, [n\log n]) = (500,3107)$. The color of a trajectory represents the community label of that agent (red represents $\mtcv_{r1}$, whereas blue represents $\mtcv_{r2}$). The dashed lines represent the initial average opinion for the two communities.}	
\end{figure}

In the first experiment, we demonstrate behavior of the expected states given in Theorems~\ref{thm_expectation} and~\ref{thm_state}~(i). Set the network size $n=500$, the range of the model $c_x = 1$, and the portion of regular agents $r_0 = 0.9$. Generate the initial value $X_i(0)$ independently from uniform distribution on $(0,1)$ for all $i \in \mtcv_{r1}$, $X_j(0)$ independently from uniform distribution on $(-1,0)$ for all $j \in \mtcv_{r2}$, and stubborn-agent states $\bfz^s$ independently from uniform distribution $(-1,1)$. In addition, let the edge weight between agents in the same community be $l_s^{(r)} = (\log^{3} n)/n$, the edge weight between communities be $l_d^{(r)} = (\log  n)/n$, and the edge weight between regular and stubborn agents be $a_{ij} = (\log n)/n$ for all $i\in \mtcv_{r}$ and $j \in \mtcv_s$. As discussed in Examples~\ref{exmp_expectation} and~\ref{exmp_state}, Fig.~\ref{fig_exp_local} shows that the agents in the same community have expected states with the same sign and close to the initial average opinion of that community, over the time interval $(n, [n\log n]) = (500,3107)$, where $[\cdot]$ is the rounding function. This experiment validates Theorem~\ref{thm_expectation} and Theorem~\ref{thm_state}~(i), which establish a correspondence between expected states and community labels.

\begin{figure}[t]
	\centering
	\includegraphics[scale=0.38]{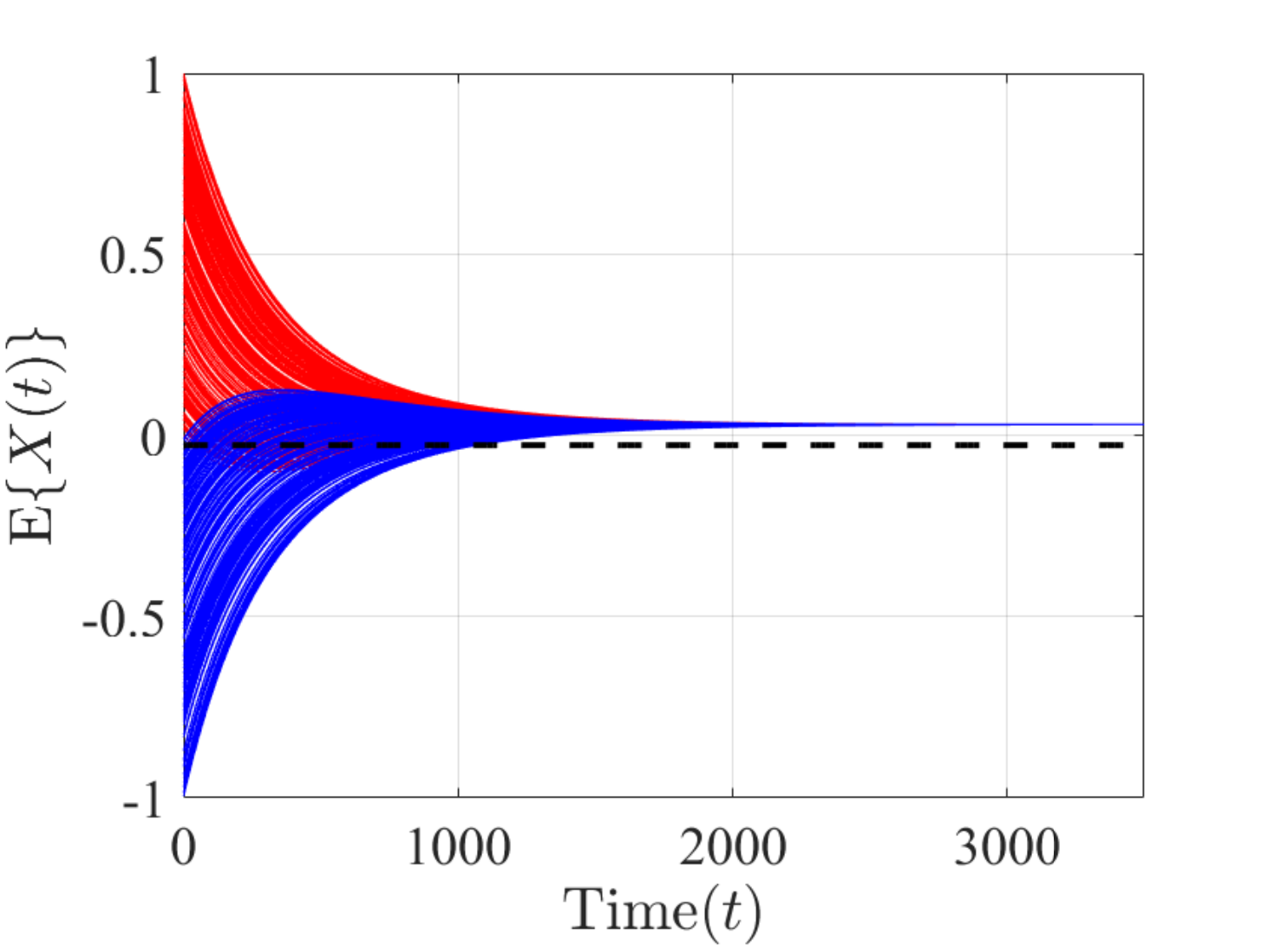}
	\caption{\label{fig_exp_global} Behavior of expected states when $l_s^{(r)} < l_d^{(r)}$. The expected states of all regular agents are close over the time interval $(n, [n\log n]) = (500,3107)$. The color of a trajectory represents the community label of that agent (red represents $\mtcv_{r1}$, whereas blue represents $\mtcv_{r2}$). The dashed line represents the initial average opinion of all regular agents.}	
\end{figure}

Now we set $l_s^{(r)} = (\log n)/n$ and $l_d^{(r)} = (\log^{3} n)/n$, and keep other parameters the same as earlier. Fig.~\ref{fig_exp_global} shows that the system is close to a global consensus, rather than a local consensus in the previous case, when connections between communities are stronger. Hence, Fig.~\ref{fig_exp_local} and~\ref{fig_exp_global} illustrate the transient behavior influenced by the community structure.

\begin{figure}[t]
	\centering
	\subfigure[\label{fig_gossip_local}The case of $l_s^{(r)} > l_d^{(r)}$.]{\includegraphics[scale=0.38]{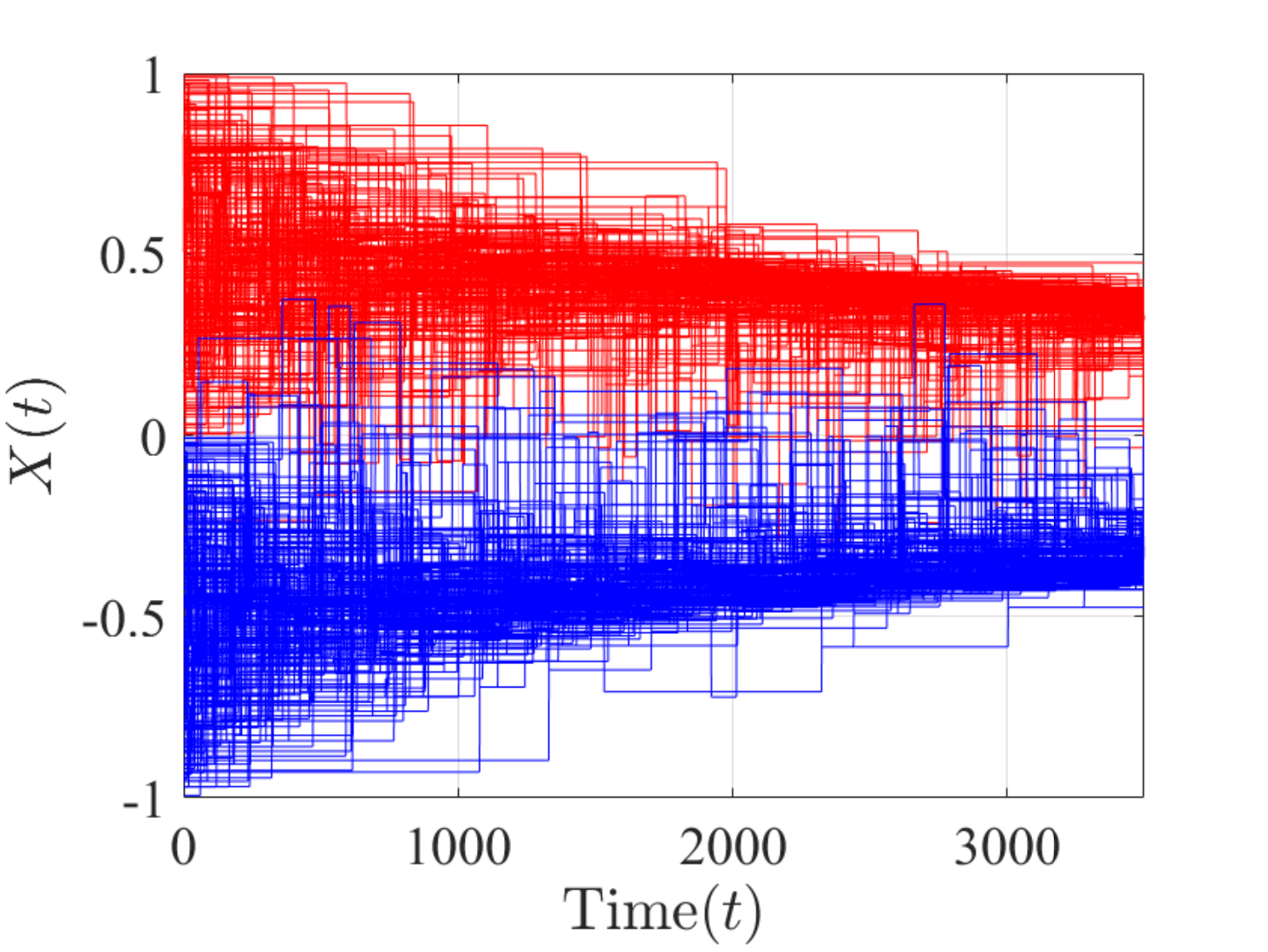}}
	\subfigure[\label{fig_gossip_global}The case of $l_s^{(r)} < l_d^{(r)}$.]{\includegraphics[scale=0.38]{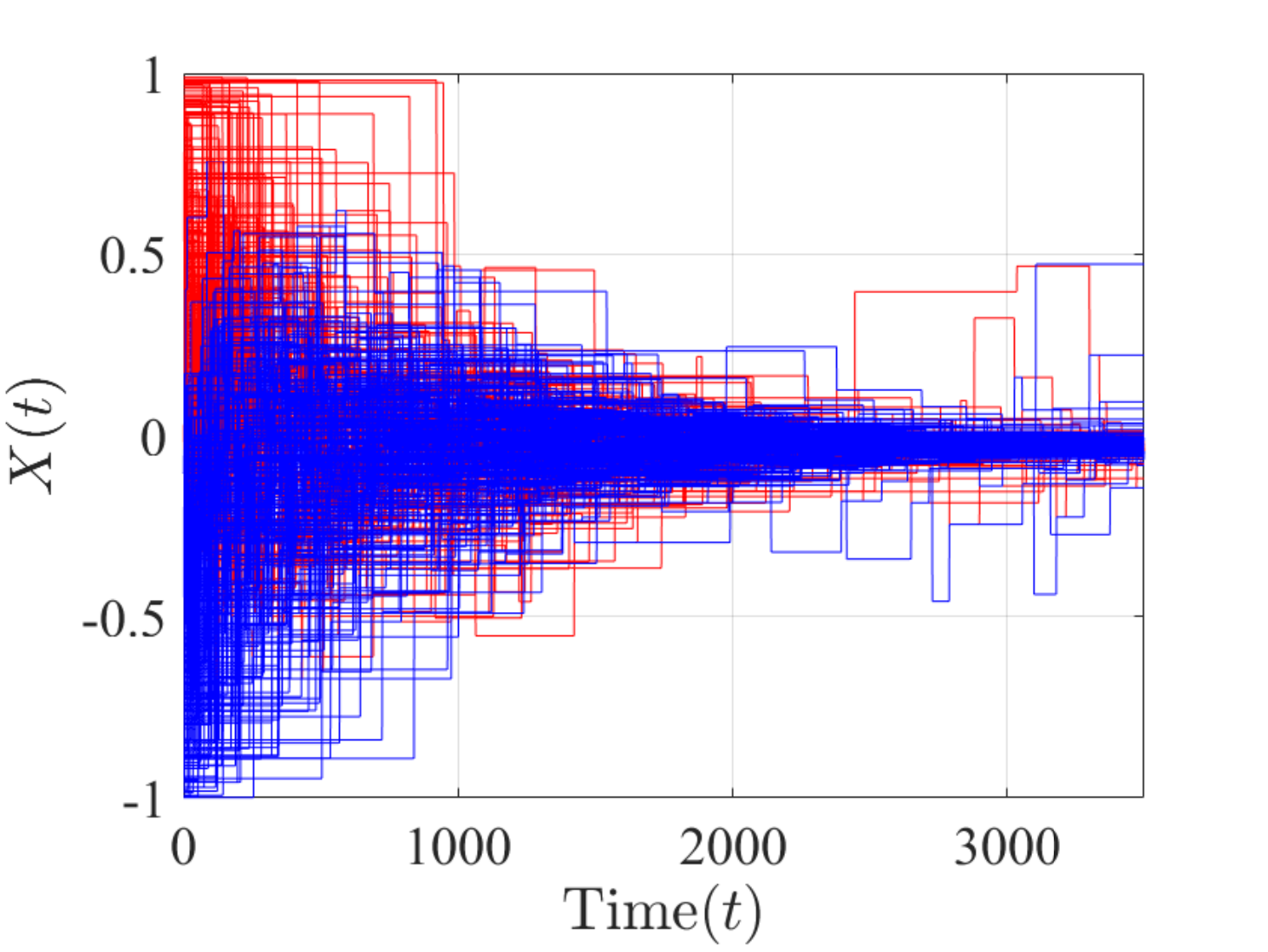}}
	\caption{\label{fig_gossip} Behavior of agent states. Behavior of agent states are similar to that of their expectations shown in Fig.~\ref{fig_exp_local} and Fig.~\ref{fig_exp_global}, over the time interval $(n, [n\log n]) = (500,3107)$. The color of a trajectory represents the community label of that agent (red represents $\mtcv_{r1}$, whereas blue represents $\mtcv_{r2}$).}
\end{figure}

In Fig.~\ref{fig_gossip} we show state evolution of the gossip model under the two previously considered  cases. As shown in Fig.~\ref{fig_gossip_local}, when connections within communities are stronger than between communities, agents states form two clusters during the transient phase, similar to their expectations. In contrast, this does not happen when connections within communities are weaker, as illustrated in Fig.~\ref{fig_gossip_global}. These observations show that the obtained results for expected states may also be valid for agent states.

\bibliographystyle{ieeetr}
\bibliography{bibliography.bib}

\end{document}